\documentclass[10pt,aps,prd,twocolumn,showpacs,superscriptaddress,nofootinbib,nobibnotes,longbibliography,floatfix]{revtex4-1} 
\usepackage[utf8]{inputenc}
\usepackage[english]{babel}
\usepackage{mathtools,amsmath,amssymb,amsfonts,eucal,mathrsfs,graphicx,tensor,csquotes,accents,commath,multirow}
\usepackage[dvipsnames]{xcolor}
\usepackage[unicode]{hyperref}
\hypersetup{colorlinks=true, citecolor=MidnightBlue,
            linkcolor=MidnightBlue, urlcolor=MidnightBlue, linktocpage=true}
\DeclareMathAlphabet{\mathpzc}{OT1}{pzc}{m}{it}
\newcommand{\note}[1]{\text{\scshape\tiny{#1}}}
\newcommand{\ee}{\mathrm{e}}

\newcommand{\dd}{\mathrm{d}}
\newcommand{\mbf}[1]{\mathbf{#1}}

\newcommand*\DAlembert{\mathop{}\!\mathbin\Box}

\newcommand{\GB}{\mathscr{G}}

\newcommand{\Ga}{\Gamma}
\newcommand{\de}{\delta}
\newcommand{\De}{\Delta}

\newcommand{\cep}{\varepsilon}
\newcommand{\ze}{\zeta}

\newcommand{\la}{\lambda}

\newcommand{\Sg}{\Sigma}

\newcommand{\Om}{\Omega}

\newcommand{\bA}{\bar{A}}
\newcommand{\bB}{\bar{B}}
\newcommand{\bph}{\bar{\phi}}
\newcommand{\bP}{\bar{P}}
\newcommand{\bQ}{\bar{Q}}
\newcommand{\bE}{\bar{E}}
\newcommand{\bC}{\bar{C}}

\newcommand{\dl}{\partial}
\newcommand{\Dl}{\nabla}

\begin{document}

\title{Fixing the dynamical evolution in scalar-Gauss-Bonnet gravity}

\author{Nicola Franchini}
\affiliation{SISSA, Via Bonomea 265, 34136 Trieste, Italy and INFN Sezione di Trieste}
\affiliation{IFPU - Institute for Fundamental Physics of the Universe, Via Beirut 2, 34014 Trieste, Italy}
\author{Miguel Bezares}
\affiliation{SISSA, Via Bonomea 265, 34136 Trieste, Italy and INFN Sezione di Trieste}
\affiliation{IFPU - Institute for Fundamental Physics of the Universe, Via Beirut 2, 34014 Trieste, Italy}
\author{Enrico Barausse}
\affiliation{SISSA, Via Bonomea 265, 34136 Trieste, Italy and INFN Sezione di Trieste}
\affiliation{IFPU - Institute for Fundamental Physics of the Universe, Via Beirut 2, 34014 Trieste, Italy}
\author{Luis Lehner}
\affiliation{Perimeter Institute for Theoretical Physics, 31 Caroline Street North, Waterloo, Ontario, N2L 2Y5, Canada}

\begin{abstract}
One of the major obstacles to testing alternative theories of gravity with gravitational-wave data from merging binaries of compact objects is the formulation of their field equations, which is often mathematically ill-suited for time evolutions. A possible way to address these delicate shortcomings is the {\it fixing-the-equations} approach, which was developed to control the behaviour of the high-frequency modes of the solutions and the potentially significant flow towards ultra-violet modes. This is particularly worrisome in gravitational collapse, where even black hole formation might be insufficient to shield regions of the spacetime where these pathologies might arise. Here, we focus (as a representative example) on scalar-Gauss-Bonnet gravity, a theory which can lead to ill-posed dynamical evolutions, but with intriguing stationary black hole physics. We study the spherical collapse of a scalar pulse to a black hole in the  {\it fixing-the-equations} approach, comparing the early stages of the evolution with the unfixed theory, and the later stages with its stationary limit. With this approach, we are able to evolve past problematic regions in the original theory, resolve black hole collapse and connect with the static black hole solutions. Our method can thus be regarded as providing a weak completion of the original theory, and the observed behaviour lends support for considering previously found black hole solutions as a natural outcome of collapse scenarios.
\end{abstract}

\maketitle

\section{Introduction}
The observation of gravitational waves emitted by binaries of compact objects opened a new possible channel to confront general relativity (GR) with alternative theories of gravity. So far, in the dynamical and strong-field regime of binary mergers, consistency with
the expectations of GR has been confirmed mostly through null and consistency tests~\cite{LIGOScientific:2016lio,LIGOScientific:2019fpa,LIGOScientific:2020tif,LIGOScientific:2021sio}.
These tests are however insufficient to rule out specific alternative theories,
as predictions for the merger waveforms of compact objects 
beyond GR have only been obtained in a handful of cases, if no approximations are made (see {\it e.g.}~\cite{Barausse:2012da,Palenzuela:2013hsa,Shibata:2013pra,Hirschmann:2017psw,Bezares:2021dma,Figueras:2021abd}). Such waveforms
require  numerically solving the field equations of a specific theory, {\it i.e.}~a full system of partial differential equations (PDEs). However, a major obstacle in this enterprise is the generic lack of well-posedness of 
the initial data (Cauchy) problem in most alternative theories of gravity~\cite{Papallo:2017qvl,Ripley:2019hxt,Ripley:2019irj,Ripley:2020vpk,Bernard:2019fjb,Bezares:2020wkn}. This problem, arising from the mathematical structure
of the underlying field equations, stands in the way of obtaining predictions of beyond-GR effects,
unless a suitable approach is introduced to address this issue.

As a representative example of this problem, we focus here on Horndeski gravity~\cite{Horndeski:1974wa}, a generalized class of scalar-tensor theories,
in which a new dynamical degree of freedom, a scalar field,  modifies GR through a non-minimal coupling. The Horndeski action is constructed so as to encompass all the possible terms leading to second-order field equations, thus avoiding Ostrogradski ghosts~\cite{Ostrogradsky:1850fid}. 
If the scalar field is invariant under constant displacements (shift symmetry), a no-hair theorem states that stationary spherically symmetric and slowly rotating black holes (BHs) are equivalent to their GR counterpart~\cite{Hui:2012qt,Sotiriou:2014pfa}. The only exception to this theorem arises  when the scalar field couples linearly\footnote{Only for a linear coupling $\phi\GB$ is the theory  shift-symmetric. In fact, the GB invariant is a total divergence, and shift symmetry becomes manifest after an integration by parts.} to the Gauss-Bonnet (GB) invariant~\cite{Sotiriou:2014pfa,Sotiriou:2013qea}, defined
in terms of the
Riemann tensor $R_{abcd}$, the Ricci tensor $R_{ab}$ and the Ricci scalar $R$ as
\begin{equation}\label{eq:GaussBonnet}
	\mathscr{G} = R_{abcd}R^{abcd} - 4 R_{ab} R^{ab} + R^2\,.
\end{equation}

When shift symmetry is broken, the phenomenology of BH solutions is even  richer. A notable example is the occurrence of spontaneous scalarization in a large sub-class of Horndeski theories~\cite{Silva:2017uqg,Doneva:2017bvd,Antoniou:2017acq,Silva:2018qhn,Cunha:2019dwb,Dima:2020yac,Herdeiro:2020wei,Berti:2020kgk,Antoniou:2021zoy,Antoniou:2022agj}. This phenomenon, like the analogous spontaneous scalarization of neutron stars in Brans-Dicke-like scalar-tensor theories~\cite{Damour:1993hw}, amounts to a growth of the scalar field around BHs, prompted by a linear tachyonic instability of the scalar and eventually quenched by the non-linearities of the problem. A tachyonic instability arises naturally in many sub-classes of Horndeski gravity~\cite{Andreou:2019ikc,Ventagli:2020rnx}, including scalar-GB (sGB) gravity, which represents the case study of this paper. 
Scalarized compact objects acquire an additional (scalar) charge, which not only affects the gravitational interaction but which can also yield monopole/dipole emission, which would impact the gravitational wave signal~\cite{PhysRevD.8.3308,Will:2005va,Barausse:2016eii}.

The appearance of intriguing physics in sGB gravity is not limited to stationary BHs, as dynamical scalarization and de-scalarization might occur  in binaries~\cite{Silva:2020omi,East:2021bqk,Doneva:2022byd,Elley:2022ept}
or in gravitational collapse~\cite{Benkel:2016kcq,Benkel:2016rlz,East:2020hgw,Corelli:2022pio,Corelli:2022phw}. However, as already stated, the well-posedness of the Cauchy problem is a necessary condition to study these time-dependent systems without approximations. The criteria of Hadamard~\cite{Hadamard10030321135} states that the Cauchy problem
for a system of PDEs is well-posed if there exists a unique solution that depends continuously on its initial data. Since proving this statement is a difficult task~\cite{evans10}, one usually attempts to show that the PDE system is
strongly hyperbolic, which is a sufficient condition for well-posedness under suitable initial and/or boundary conditions~\cite{Hilditch:2013sba,Sarbach:2012pr}. This requires 
writing the system of equations as a quasilinear first-order system and showing that the principal part has real eigenvalues and a complete set of eigenvectors. 

For sufficiently weak couplings and smooth initial data, local well-posedness of sGB gravity can be established~\cite{Kovacs:2020pns,Kovacs:2020ywu} and successful studies have been presented in this regime~\cite{Ripley:2019hxt,Ripley:2019irj,Ripley:2020vpk}. 
However, for stronger couplings and/or scenarios where higher frequency modes develop, the hyperbolic structure of the system can break down. Specifically, it can be shown that the underlying equations can switch their character  from hyperbolic to elliptic~\cite{Ripley:2019hxt,Ripley:2019irj,East:2021bqk}. More specifically, during the evolution one of the characteristic speeds (i.e. one of the eigenvalues of the characteristic matrix) becomes imaginary. This breakdown of the character of the system resembles the Tricomi equation~\cite{Ripley:2019hxt,Ripley:2019irj,Ripley:2020vpk}.

Faced with this possibility, a potential way through is to modify the system of equations
so that short wavelength features (with respect to some chosen scale) 
in the solution are somewhat controlled, while longer wavelength modes are faithfully tracked~\cite{Cayuso:2017iqc}.
This idea, applied successfully in many studies~\cite{Allwright:2018rut,Cayuso:2020lca,Bezares:2021yek,Lara:2021piy,Gerhardinger:2022bcw}, is rooted in the expectations that the theory at hand is the truncation of a 
putative theory which has a sensible behavior, and that the problems encountered are an artifact of
truncating it at a given order.  As an analogy, consider the pinch-off of a soap 
bubble. Some time prior to and after the pinch-off, its surface can be faithfully described with hydrodynamics, but
around the pinch-off itself one needs to use a different theory (molecular dynamics)  or a suitable effective regularization of the hydrodynamics equations to bridge the earlier and 
later stages. If the solution obtained is largely insensitive to the details of
the modification introduced and, stronger yet, such solution recovers the characteristics
expected in the early and late stages of the dynamics, strong support for the approach yielding a faithful 
representation of the UV complete underlying theory can be argued.

In the current work, we focus on a problem that illustrates this approach and behavior.
Namely, we study the collapse within sGB and show that the latter gives rise to a scalarized BH after passing through a regime where the original system of field equations develops a pathological behavior.
We implement a technique that controls analytical obstacles and related numerically 
induced problems in the time evolution, and show that this approach
allows for bridging regimes where the theory is sensible. This latter property --- not necessarily
expected {\it a priori} --- is undoubtedly a welcome sight that indicates the robustness of scalarized BHs as fixed point solutions of the theory.

This work is organized as follows. In section~\ref{sec:theory} we present the theory of sGB, its equations of motion, and how we replace those equations with the fixed system. Then, in section~\ref{sec:spherical_symmetry}, we specialize to spherical symmetry, expressing the equations of motion in the {\it full} and in the {\it fixed} theory,
and presenting the characteristic speeds. Moreover, we discuss the numerical set up of the simulations. The results of the latter are shown in section~\ref{sec:results}. Finally, we draw our conclusions and possible implications of our results in section~\ref{sec:conlusions}. By convention, we use units such that $c=G_\note{N}=1$.

\section{The theory and the fixing}\label{sec:theory}

The action for sGB gravity in vacuum (also referred to as {\it full} theory henceforth) is given by~\cite{Julie:2019sab}
\begin{align}
	S_\note{GB} = \int\frac{\dd^4 x \sqrt{-g}}{16\pi} \Bigg[ R -\frac{1}{2}\left(\dl_a\phi\right)^2 + f(\phi)\mathscr{G} \Bigg]\,,
\end{align}
where the GB invariant $\GB$ given in Eq.~\eqref{eq:GaussBonnet} couples to the scalar field through the function $f(\phi)$.
A variation of the action gives the equations of motion
\begin{align}
	\label{eq:Einstein}
	& R_{ab} - \frac{1}{2}g_{ab}R = T_{ab}^{(\phi)} + T_{ab}^{\left(\mathscr{G}\right)} \,, \\
	\label{eq:KleinGordon}
	& \DAlembert \phi = S^{(\GB)} \,,
\end{align}
where the sources of the Einstein and Klein-Gordon equations are
\begin{align}
	T_{ab}^{(\phi)} & = \frac{1}{2}\dl_a\phi\,\dl_b\phi -\frac{1}{4}g_{ab}\left(\dl_c\phi\right)^2 \,, \\
	T_{ab}^{\left(\mathscr{G}\right)} & = - 4 P_{acbd} \Dl^c\Dl^d f \,, \qquad S^{(\GB)} = - f'(\phi) \mathscr{G}
\end{align}
with
\begin{equation}
    P_{abcd} = R_{abcd} - 2 g_{a\left[c\right.} R_{\left.d\right] b} + 2 g_{b \left[c\right.} R_{\left.d\right] a} + g_{a\left[c\right.}g_{\left.d\right] b} R \,.
\end{equation}

As discussed in the Introduction, the system~\eqref{eq:Einstein}--\eqref{eq:KleinGordon} is not strongly hyperbolic for a large class of choices of the coupling function $f(\phi)$. In order to circumvent this problem we introduce here a {\it fixing-the-equations} technique~\cite{Cayuso:2017iqc}, which was inspired by dissipative relativistic hydrodynamics~\cite{1976PhLA...58..213I,Israel:1976tn,Muller:1967zza}. The main idea is to modify in an {\it ad hoc} way the higher-order contributions to the equations of motion, by replacing them with some auxiliary fields, and let the latter relax towards their correct value through a driver equation.

We replace the system~\eqref{eq:Einstein}-\eqref{eq:KleinGordon} with the following {\it fixed} system of equations.
\begin{align}
	\label{eq:IS_Einstein}
	& R_{ab}-\frac{1}{2} g_{ab} R = T_{ab}^{(\phi)} + \Gamma_{ab} \,, \\
	\label{eq:IS_KG}
	& \DAlembert\phi = \Sg \,, \\
	\label{eq:IS_Aux}
	& \xi\DAlembert \mbf{u} + \tau\dl_t \mbf{u} - \left( \mbf{u} - \mbf{S} \right) = 0\,.
\end{align}
Here, we gathered the auxiliary fields in the vector $\mbf{u} = \left( \Ga_{ab}, \Sg \right)$, while $\{\xi,\tau\}$ are constant timescales (although in principle they can differ for each component of $\mbf{u}$)
controlling how the auxiliary variables approach to the full theory solution $\mbf{S} = \left( T_{ab}^{(\GB)}, S^{(\GB)} \right)$. Notice also that equation~\eqref{eq:IS_Aux} is one of many options
that can force the auxiliary variables to approach their corresponding correct values.
{Crucial in this respect is the presence of the first time derivatives $\partial_t$, which introduce a ``preferred'' time direction, {\it i.e.}, dissipation.}
The choice, made above, of a wave-like driver equation for the auxiliary fields
 is in a sense natural given the underlying hyperbolic structure
of the Einstein equations. We will discuss some issues related to this choice later.
In the following, we focus our analysis and numerical evolution on spherical symmetry.

\section{Evolution in Spherical symmetry}\label{sec:spherical_symmetry}
In order to study the non-linear dynamics we write down the equations \eqref{eq:Einstein}--\eqref{eq:KleinGordon} or \eqref{eq:IS_Einstein}--\eqref{eq:IS_Aux} as a first-order
(in time) evolution system, by applying the techniques used in Refs.~\cite{Bernard:2019fjb,Ripley:2019hxt}.

\subsection{Evolution equations}
We adopt the following ansatz for the metric in  polar coordinates
\begin{equation}
\label{eq:polar}
	g_{ab} \dd x^a \dd x^b = -\ee^{2A(t,r)} \dd t^2 + \ee^{2B(t,r)} \dd r^2 + r^2 \dd \Om^2\,,
\end{equation}
and we introduce two new functions to express derivatives of the scalar field---which is a generic function of time and radial coordinate $\phi = \phi(t,r)$---by defining
\begin{equation}
    \label{eq:defPQ}
	P(t,r) \equiv \ee^{-A+B} \dl_t \phi \,, \qquad Q(t,r) \equiv \dl_r \phi\,.
\end{equation}
In spherical symmetry, the tensor $T_{ab}^{(\GB)}$ has four non-trivial components. This means that in the {\it fixed} theory we make use of five auxiliary variables, $\Sg(t,r)$, $\Ga_{11}(t,r)$, $\Ga_{12}(t,r)$, $\Ga_{22}(t,r)$ and $\Ga_{33}(t,r)$. Henceforth, we place a bar on top of any function evaluated in the {\it full} theory, in order to distinguish it from the {\it fixed} theory.

Schematically, the {\it full} system~\eqref{eq:Einstein}--\eqref{eq:KleinGordon}, written in spherical symmetry with the ansatz considered and after some manipulation of the equations, reduces to three evolution equations and two constraint equations
\begin{align}
	\label{eq:Eqphi}
	& \bE_\phi^\note{GB} \equiv \dl_t \bph - \ee^{\bA-\bB} \bP = 0 \,,\\ \label{eq:EqQ}
	& \bE_Q^\note{GB} \equiv \dl_t \bQ - \dl_r\left(\ee^{\bA-\bB} \bP\right) = 0 \,,\\ 
	\label{eq:EqP}
	& \bE_P^\note{GB} \equiv \bE_P^\note{GB}\big(\dl_t \bP;\bA,\bB,\bph, \bP, \dl_r \bP, \bQ, \dl_r \bQ\big) = 0\,,\\
	\label{eq:EqA}\
	& \bC_A^\note{GB} \equiv \bC_A^\note{GB}\big(\dl_r \bA;\bB, \bph, \bP, \dl_r \bP, \bQ, \dl_r \bQ\big) = 0\,, \\
	\label{eq:Beq}
	& \bC_B^\note{GB} \equiv \bC_B^\note{GB}\big(\dl_r \bB;\bB, \bph, \bP, \dl_r \bP, \bQ, \dl_r \bQ\big)=0\,,
\end{align}
where $\bE_P^\note{GB}$, $\bC_A^\note{GB}$, $\bC_B^\note{GB}$ have a lengthy and uninformative expression that can be found in the appendix of reference~\cite{Ripley:2019irj}. The first two equations come from the consistency of the partial derivatives of the scalar field, the third  comes from the Klein-Gordon equation~\eqref{eq:KleinGordon} and the last two  come from the constraints of the theory~\eqref{eq:Einstein}.

On the other hand, the evolution equations for the {\it fixed} theory in spherical symmetry are explicitly given by
\begin{align}
    \label{eq:phifix}
    E_\phi^\note{F} \equiv & \, \dl_t \phi - \ee^{A-B} P = 0\,, \\
    \label{eq:Qfix}
    E_Q^\note{F} \equiv & \, \dl_t Q - \dl_r\left(\ee^{A-B} P\right) = 0\,, \\
    E_P^\note{F} \equiv & \, \dl_t P - \frac{1}{r^2} \dl_r \left( r^2 \ee^{A-B} Q \right)  + \ee^{A+B} \Sg = 0 \,, \\
    \label{eq:ufix}
    E_\mbf{u}^\note{F} \equiv & \, \dl_t \mbf{u} - \ee^{A-B} \mbf{H} = 0   \,,  \\
    \label{eq:Jfix}
    E_\mbf{J}^\note{F} \equiv & \, \dl_t \mbf{J} - \dl_r\left(\ee^{A-B} \mbf{H}\right)  = 0  \,,  \\
    E_\mbf{H}^\note{F} \equiv & \, \dl_t \mbf{H} - \frac{1}{r^2} \dl_r \left( r^2 \ee^{A-B} \mbf{J} \right) - \frac{\tau}{\xi}\ee^{2A}\mbf{H} \notag \\
    & - \frac{1}{\xi}\ee^{A-B} \left(\mbf{S} - \mbf{u}\right) = 0 \,, \\
    \label{eq:Afix}
    C_A^\note{F} \equiv & \, \dl_r A - \frac{r}{8}  \left(P^2+Q^2 + 4 \Gamma _{22}\right) \notag \\
    & + \frac{1-\ee^{2 B}}{2 r} = 0 \,, \\
    \label{eq:Bfix}
    C_B^\note{F} \equiv & \, \dl_rB - \frac{r}{8} \left(P^2+Q^2 + 4 \Gamma _{11} \ee^{-2(A-B)}\right) \notag \\
    & - \frac{1-\ee^{2 B}}{2 r} = 0 \,,
\end{align}
where $\mbf{H} \equiv \ee^{-A+B} \dl_t\mbf{u}$ and $\mbf{J} \equiv \dl_r\mbf{u}$. The explicit form for $\mbf{S}$ is given in Appendix~\ref{app:source}. 

The two systems we are considering are described by two constraint equations and $M$ evolution equations, where $M=3$ for the {\it full} theory and $M = 3 + 3\cdot5 = 18$ for the {\it fixed} theory. If we label with $\mbf{V}$ the variables corresponding to the evolution equations (with dimension $M$) and with $\mbf{W}=\left(A,B\right)$ the two metric variables, we can write the system in the following compact form
\begin{align}
    & E_I\big(\dl_t \mbf{V}, \dl_r \mbf{V}, \mbf{V}, \dl_r \mbf{W}, \mbf{W} \big)=0\,, \\ & C_L\big(\dl_r \mbf{V}, \mbf{V}, \dl_r \mbf{W}, \mbf{W} \big)=0\,.
\end{align}
An overview of the variables and of the equations can be found in Table~\ref{tab:functions}.

\begin{table}
\begin{tabular}{p{0.22\columnwidth}|p{0.25\columnwidth}|p{0.45\columnwidth}}
                        & Full theory & Fixed theory \\
 \hline
 Variables $\mbf{V}$    & $\bph$, $\bP$, $\bQ$ & $\phi$, $P$, $Q$, $\mbf{u}$, $\mbf{H}$, $\mbf{J}$ \\
 Variables $\mbf{W}$    & $\bA$, $\bB$ & $A$, $B$  \\
 \hline
 Evolution equations    & $\bE_\phi^\note{GB}$, $\bE_Q^\note{GB}$, $\bE_P^\note{GB}$ & $E_\phi^\note{F}$, $E_Q^\note{F}$, $E_P^\note{F}$, $E_\mbf{u}^\note{F}$, $E_\mbf{J}^\note{F}$, $E_\mbf{H}^\note{F}$ \\
 Constraints            & $\bC_A^\note{GB}$, $\bC_B^\note{GB}$ & $C_A^\note{F}$, $C_B^\note{F}$
\end{tabular}
\caption{List of the variables used and characteristic speeds computed in the {\it full} and in the {\it fixed} theories.}\label{tab:functions}
\end{table}

A useful tool to check the hyperbolicity of the system during the evolution is to evaluate the characteristic speeds. The latter can be calculated from the characteristic matrix of the full system
\begin{equation}
    \mathpzc{P}(\xi) = 
    \begin{pmatrix}
    \mathpzc{A} \chi_t + \mathpzc{B} \chi_r & \mathpzc{Q} \chi_r \\
    \mathpzc{R} \chi_r & \mathpzc{S} \chi_r
    \end{pmatrix}\,,
\end{equation}
where we introduced the covector $\chi_a$ and defined
\begin{align}
    \mathpzc{A}^{IJ} = & \frac{\de E_J}{\de (\dl_t V_I)} \,, \quad \mathpzc{B}^{IJ} = \frac{\de E_J}{\de (\dl_r V_I)}\,, \\
    \mathpzc{Q}^{IJ} = & \frac{\de E_J}{\de (\dl_r W_I)} \,,  \quad \mathpzc{R}^{IJ} = \frac{\de C_J}{\de (\dl_r V_I)} \,, \\
    \mathpzc{S}^{IJ} = & \frac{\de C_J}{\de (\dl_r W_I)} \,,
\end{align}
with the indices $I$ and $J$ running from $1$ to either $2$ or $M$ according to the function and the equation considered.
The characteristic speeds are then defined as
\begin{align}
    v_\pm = & \,\frac{1}{2} \left( \mathrm{Tr}(\mathpzc{C}) \pm \sqrt{\mathpzc{D} } \right) \,, \\
    \mathpzc{D} = & \,\mathrm{Tr}(\mathpzc{C})^2 - 4\mathrm{Det}(\mathpzc{C}) \,,
\end{align}
where, if $\mathpzc{S}$ is invertible,
\begin{equation}
    \mathpzc{C} \equiv \mathpzc{A}^{-1} \cdot \left( \mathpzc{B} - \mathpzc{Q} \cdot \mathpzc{S}^{-1} \cdot \mathpzc{R} \right) \,.
\end{equation}
The expression for the characteristic speeds for the {\it full} theory $\bar{v}_\pm^\note{GB}$ is very long and we do not show it here. On the other hand, the speeds in the {\it fixed} theory are equivalent to those of GR
\begin{equation}
    v_\pm^\note{F} = \pm \ee^{A-B}\,.
\end{equation}

For the rest of the paper, we choose the coupling function
\begin{equation}
    f(\phi) \equiv \frac{1}{8}\eta\phi^2 + \frac{1}{16}\ze\phi^4\,,
\end{equation}
where $\eta$ and $\ze$ are parameters of the theory. The quadratic term proportional to $\eta$ is the one responsible for spontaneous scalarization provided that $\eta > 0$. The addition of a quartic term proportional to $\ze$ ensures the existence of radially stable spherically symmetric BH solutions, provided that $\ze \lesssim - 0.8 \eta$~\cite{Silva:2018qhn}. In all the cases considered we choose $\ze = -6\eta$, which largely ensures this requirement. The hyperbolicity of these class of theories was largely studied in previous works. The ansatz we chose for spherically symmetric evolution in polar coordinates leads to a breaking of the hyperbolicity in the linear theory~\cite{Ripley:2019irj,Ripley:2019hxt}, and we confirmed that it also happens for this choice of the coupling (see {\it e.g.}, Figure~\ref{fig:speeds} in the Results section). Attempts to evolve these theories were made with the help of excision~\cite{Ripley:2020vpk}
and/or generalised harmonic gauge~\cite{Kovacs:2020pns,Kovacs:2020ywu,East:2021bqk}. Both help to
ensure a (local) well-posed formulation of the problem by adopting a convenient gauge as well as
excising a significant region inside the BH, where short wavelength modes are excited.
However, even with these techniques, there exist regions inside whose the system dynamically develops elliptic character, thus
breaking well-posedness. The approach taken with the {\it fixing-the-equations} method aims
to provide a robust way to explore a potentially significantly larger region of the parameter space.

\subsection{Diagnostic tools}
For diagnostic of the solutions obtained we find it useful to define the following norm, valid for any quantity $F$ with two different definitions in the {\it full} and {\it fixed} theory
\begin{equation}
    ||F|| = \frac{|F^\note{F} - F^\note{GB}|}{|F^\note{F}| + |F^\note{GB}| + \cep}\,,
\end{equation}
where $\cep = 10^{-5}$ is to ensure that the denominator never goes to zero.
We stress that since the dynamical evolution cannot be followed in the {\it full} theory past the formation of the elliptic region, in our analysis we often compute the quantities $F^\note{GB}$ using the variables of the {\it fixed} theory.

From the metric at infinity we can define the Misner-Sharp mass~\cite{Misner:1964je} as
\begin{equation}\label{MSmass}
M_\note{MS}(t) = \left.\frac{r}{2}\left(1 - \ee^{-2B(t,r)}\right)\right|_{r=r_\note{max}}\,.
\end{equation}
During the evolution, we rescale all the dimensionful quantities such as $M_\note{MS}(0) = M \rightarrow 1$. For example, we redefine $r\rightarrow r/M$ and $\eta \rightarrow \eta M^2$.

Finally, we compute the null convergence condition (NCC) as $R_{ab}\ell^a\ell^b\geq 0$, where the contraction of the Ricci tensor is taken with null vectors defined as $\ell^a = \{\ee^{-A},\ee^{-B},0,0\}$. A satisfied NCC is expected to hold in GR for BHs and cosmological settings, and it is a powerful tool in many theorems~\cite{Bardeen:1973gs,Hayward:1993wb}. We expect anyway that the NCC is violated in sGB gravity~\cite{Ripley:2019irj}. It is worth noting that since $R_{ab}$ contains time derivatives, one can get rid of them using the equations of motion. This means that the NCC has a different expression when evaluated in the {\it full} or in the {\it fixed} system.

\subsection{Initial data}
As initial data (ID) for $\phi$, $P$ and $Q$, we take two different families of function. The ID labelled as type I is a static Gaussian pulse
\begin{eqnarray}
\label{eq:initialdata1}
	\phi(0,r) &=& a_0 \exp\left[-\left(\frac{r-r_0}{w_0}\right)^2\right]\,,\\
    Q(0,r) &=& \dl_r \phi(0,r)~,\\
    P(0,r) &=& 0~,
\end{eqnarray}
where $a_{0}$ and $r_{0}$ are the amplitude and grid location of the pulse respectively, and $w_0$ is the root-mean-square width of the Gaussian pulse. The ID labelled as type II is an (approximately) ingoing pulse, defined as
\begin{eqnarray}
\label{id:in1}
\phi(0,r) &=& a_0 \left(\frac{r}{w_0}\right)^2~ \exp\left[-\left(\frac{r-r_0}{w_0}\right)^2\right]\,,\\
\label{id:in2}
Q(0,r) &=& \dl_r \phi(0,r)\,,\\
\label{id:in3}
P(0,r) &=& -\frac{1}{r} \phi(0,r) - Q(0,r)\,.
\end{eqnarray}
For the metric variables $A$ and $B$ we do not impose any initial data, as one can just solve the constraints equations given the initial configuration for the scalar, as explained below.
Finally, we set all the auxiliary variables to $0$ at $t=0$. 

\subsection{Numerical scheme}
We solve the two systems of equations with the same methodology, by using a fully constrained evolution scheme which works as follows.

We discretize all the equations over the domain $r \in \left[0,r_\note{max}\right]$. By choosing a resolution $N$ we determine the radial step $\De r = r_\note{max}/N$. To define the time step $\De t$ we adopt a Courant parameter $\la_r=0.25$ such that $\De t= \la_{r} \De r $ satisfies the Courant-Friedrichs-Levy condition. To ensure the regularity of the equations at the origin, we must impose
\begin{align}\label{eq:BC_metric}
	& \dl_r A(t,0) = 0 \,, \qquad B(t,0) = 0\,, \qquad \dl_r B(t,0) = 0 \,, \\
	& \dl_r P(t,0) = 0 \,, \qquad Q(t,0) = 0 \,.
\end{align}
These conditions apply both to the {\it full} and the {\it fixed} system. Moreover, for the latter, we also set the following conditions for the auxiliary variables
\begin{equation}
    \dl_r \mbf{u}(t,0) = 0 \,, \qquad \dl_r \mbf{H}(t,0) = 0 \,, \qquad \mbf{J}(t,0) = 0 \,.
\end{equation}
At the outer boundary of integration, we set approximately outgoing boundary conditions for the variables $\mbf{V}$, while the metric variables are automatically determined by the solution of the equations.

We perform the evolution as follows. At each time step, we solve the two first order equations $C_A$ and $C_B$ from $0$ to $r_\note{max}$ in space by using a fourth-order Runge-Kutta (RK) scheme to obtain a solution for $A$ and $B$. As to carry the integration one needs to know the value of the functions $\mbf{V}$ at intermediate \textit{virtual} points, we evaluated this with fifth-order Lagrangian interpolator. At $r=0$, it is sufficient to set $A(t,0)=0$ and $B(t,0)=0$ to automatically satisfy conditions~\eqref{eq:BC_metric}. Once a solution is found, we exploit the remaining gauge freedom to set $A(t,r) \rightarrow A(t,r) - A(t,r_\note{max})$. This ensures that the proper time of an observer located at $r=r_\note{max}$ is identically $t$.

Once the metric functions are obtained, we integrate the other variables $\mbf{V}$ in time through the method of lines by using a fourth-order accurate strong stability preserving RK integrator. For convenience, we also added sixth-order Kreiss-Olliger dissipation. In every equation, we discretized the radial derivatives using fourth-order finite differences operators satisfying summation by parts~\cite{Calabrese:2003vx}.

\section{Results}\label{sec:results}
In this section, we show how the {\it fixing-the-equation} method  allows for a stable evolution in the regime where the {\it full} theory would break down due to the appearance of an elliptic region. A summary of the numbered evolutions (EV\#) studied in this Section is shown in Table~\ref{tab:runs}.

\begin{table}
    \centering
    \begin{tabular}{c|c|cccc|c|cc|c}
       \multirow{2}*{Label} & \multirow{2}*{Theory} & \multicolumn{4}{c|}{Initial Data} & \multirow{2}*{$\,\eta\,$} & \multirow{2}*{$\xi/\eta^2$} & \multirow{2}*{$\tau/\eta$} & \multirow{2}*{Figure} \\
       &                       & type & $a_0$ & $r_0$ & $w_0$   &                        &  & &   \\
       \hline
        EV1 & sGB   & I & 0.1 & 25 & 6 & $6$ & \textbackslash & \textbackslash & \multirow{2}*{Fig.~\ref{fig:speeds}} \\
        EV2 & Fixed & I & 0.1 & 25 & 6 & $6$ & 0.01 & 0 &  \\
        \hline
        EV3 & Fixed & II & 0.03 & 25 & 6 & $6$ & $1$ & 0 & Figs.~\ref{fig:rhsA}--\ref{fig:conv} \\
        \hline
        EV4 & GR & II & 0.03 & 25 & 6 & \textbackslash & \textbackslash & \textbackslash & Fig.~\ref{fig:NCC} \\
        \hline
        EV5 & Fixed & II & 0.03 & 25 & 6 & $8$ & 1 & 0 & Fig.~\ref{fig:phi_e6z6} \\
        \hline
        EV6 & Fixed & II & 0.03 & 25 & 6 & $6$ & $2$ & 0 & \multirow{2}*{Fig.~\ref{fig:extrapolation}} \\
        EV7 & Fixed & II & 0.03 & 25 & 6 & $6$ & $3$ & 0 & \\
        \hline
        EV8 & Fixed & II & $10^{-3}$ & 25 & 6 & $6$ & $1$ & 0 & Fig.~\ref{fig:conv}
    \end{tabular}
    \caption{Summary of the runs made, where we specify an identifying label, in which theory we are solving  the equations, the kind of initial data, the parameters of the theory, the parameters of the fixing and in which figure they are used. The label sGB indicates the {\it full} theory, while the label GR indicates a  theory with minimally coupled scalar field.}
    \label{tab:runs}
\end{table}

\begin{figure}
	\centering
	\includegraphics[width=\columnwidth]{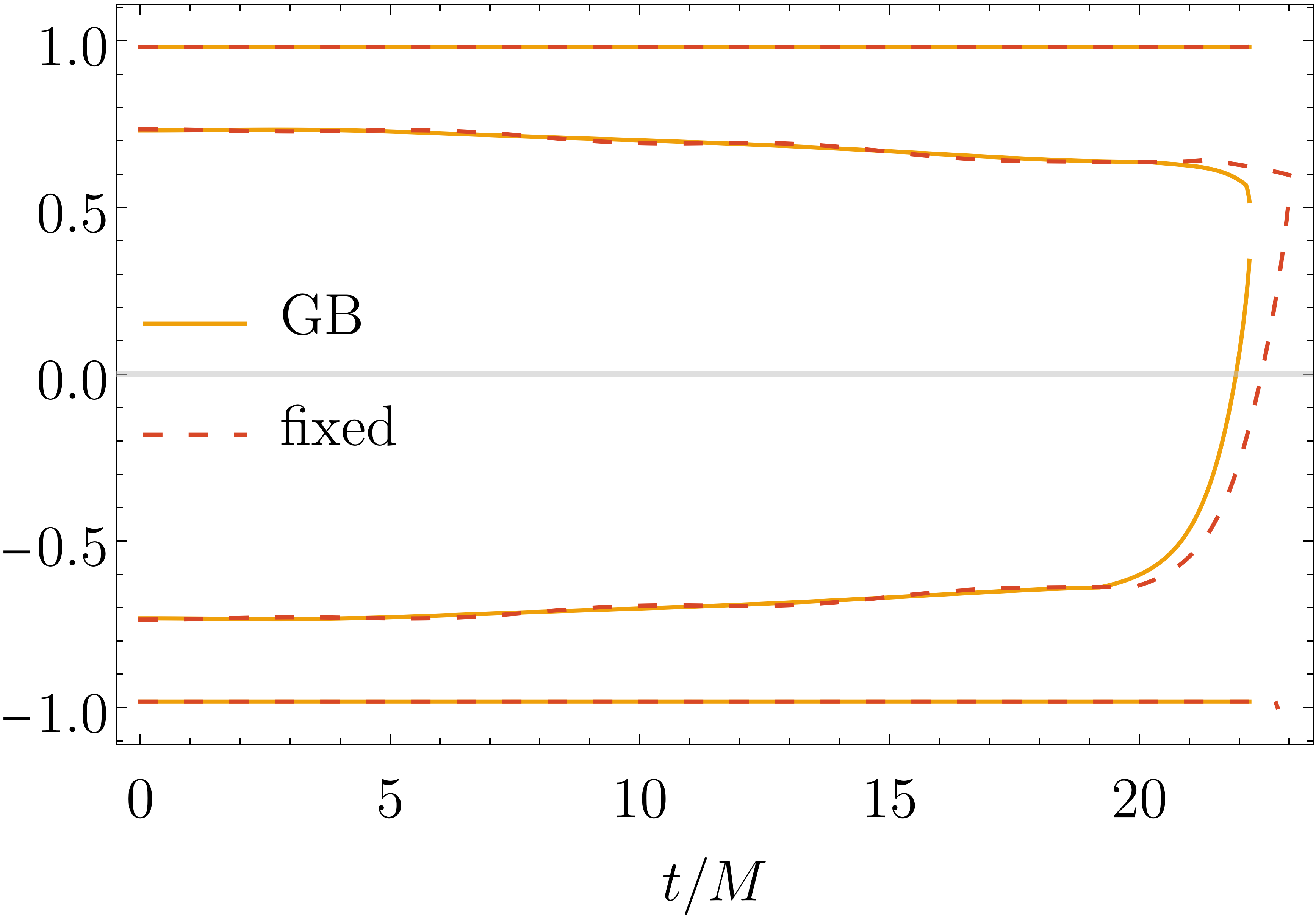}
	\caption{A comparison of the maximum and the minimum of $\bar{v}^\note{GB}_\pm$ and $v^\note{GB}_\pm$, respectively represented by a yellow solid line and a red dashed line. From top to bottom, we show $\max(v_+)$, $\min(v_+)$, $\max(v_-)$, $\min(v_-)$.} \label{fig:speeds}
\end{figure}

First of all, we show that  gravitational collapse in the {\it full} and {\it fixed} theories produces approximately the same dynamics. We evolve the same pulse in the two theories with the choice of parameters summarized in Table~\ref{tab:runs}, with labels EV1 and EV2. In the comparison, we pay attention to the characteristic speeds of the theories. In Figure~\ref{fig:speeds} we plot, as a function of time, the maximum and minimum of $\bar{v}^\note{GB}_\pm$ and ${v}^\note{GB}_\pm$. We stress that the functional form of the two quantities is the same, and a bar (no bar) denotes quantities evaluated in the {\it full} ({\it fixed}) theory. During the first stage of the gravitational collapse, around $t\leq15 M$, the characteristic speeds are almost indistinguishable and constant during the evolution, showing that the character of the {\it full} theory is strongly hyperbolic and well reproduced by the {\it fixed} one. Nevertheless, at $t\sim 22 M,$ one of the characteristic speeds crosses zero becoming imaginary shortly after, and therefore the system changes character from hyperbolic to elliptic. It is worth noting that even if the equations in the {\it fixed} system are different, their variables reproduce quite well the breaking of hyperbolicity of the {\it full} theory. We recall that the true characteristic speeds of the {\it fixed} problem are $v^\note{F}_\pm$, and that they never change sign. This means that for every case, for $t \gtrsim 22M$ the {\it full} theory cannot be evolved, but we can always follow the evolution of the {\it fixed} system.

\begin{figure*}
	\centering
	\includegraphics[width=\textwidth]{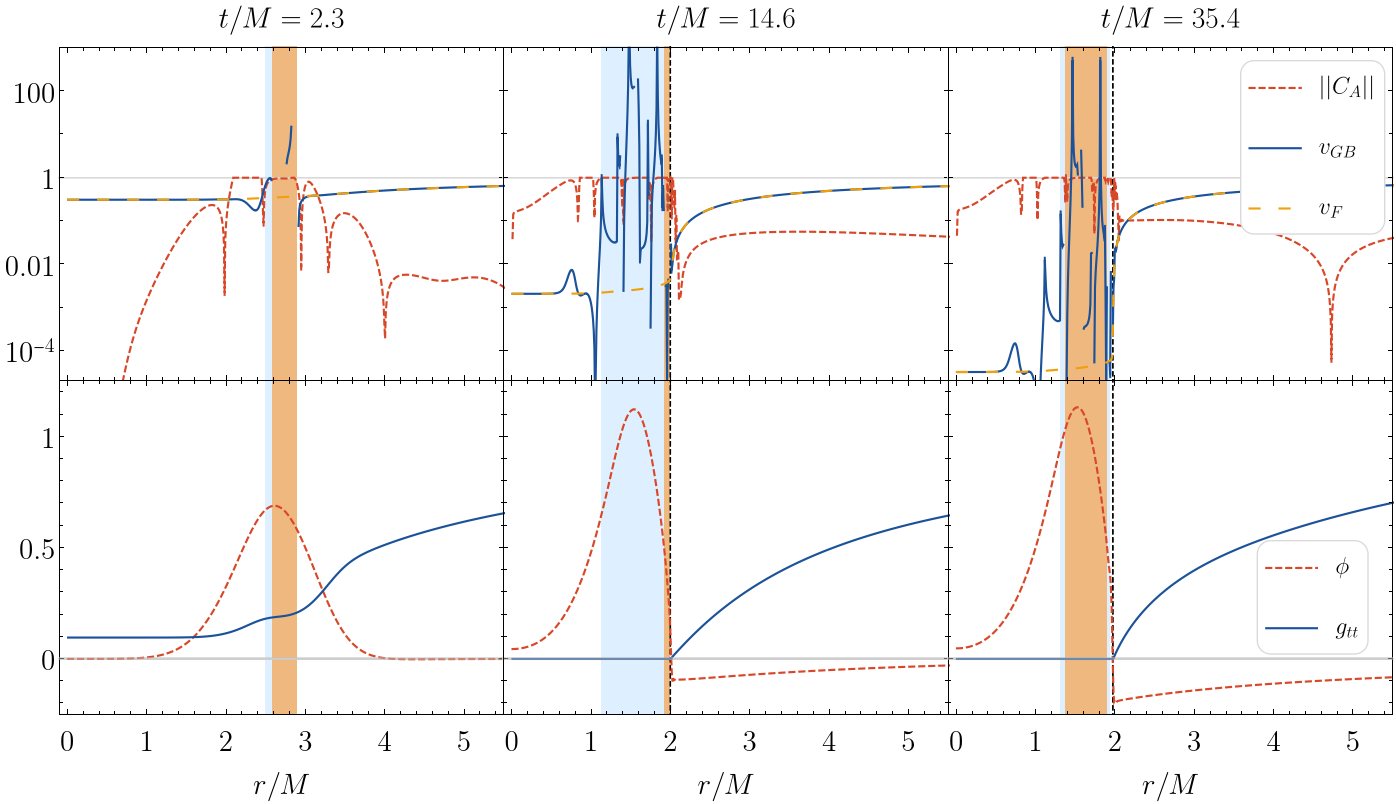}
	\caption{Top panel: snapshots of characteristic speeds $v_+^\note{GB}$ and $v_+^\note{F}$ (blue solid and yellow long-dashed lines respectively) compared to the relative difference between $C_A^\note{GB}$ and $C_A^\note{F}$ (red short-dashed line). 
	Bottom panel: scalar field $\phi$ (red dashed line) and lapse $\ee^{A}$ (blue solid line).
	The first snapshot is  when the elliptic region forms in the full theory. The second snapshot is when the apparent horizon forms in the fixed theory. The third snapshot is at a late time when the scalarized profile appeared in the fixed theory. The light blue area corresponds to the region where $v_+^\note{GB}<0$ or $v_+^\note{GB}>1$, the orange area corresponds to the elliptic region (in the {\it full} system) where $\mathpzc{D}<0$, while the black dashed vertical line in the second and third panels is the location of the apparent horizon.} \label{fig:rhsA}
\end{figure*}

With this in mind, we evolve a pulse in the {\it fixed} theory, for a choice of parameters that would break hyperbolicity in the {\it full} theory. In Figure~\ref{fig:rhsA}, we give an estimate of how the {\it full} theory would behave, by computing the quantities of this theory with the variables of the {\it fixed} one. Here, we plot time snapshots of the evolution EV3, summarized in Table~\eqref{tab:runs}. In the top panels, we show the characteristic speeds, $v_\pm^\note{GB}$ and $v_\pm^\note{F}$, and the norm $||C_A||$. In the bottom panels we show the scalar profile $\phi$ and $g_{tt}$. The first panel shows results at $t = 2.3M$. One can see that the {\it full} system would develop an elliptic character, confined only in a limited region of space where the characteristic speeds would become imaginary and $\mathpzc{D}<0$. This region is denoted by an orange vertical strip. Moreover, we plot a blue region defined  by the conditions $v_\pm^\note{GB}<0$ or $v_\pm^\note{GB}>1$. This region roughly corresponds to that where the deviation between $C_A^\note{F}$ and $C_A^\note{GB}$ becomes noticeable. This pattern remains in the second and third panels, the former being a snapshot at BH formation  at $t \simeq 14.6M$, while the latter is a snapshot of the later evolution at $t = 35.4M$, where we expect scalarization of the BH. 

It is worth noticing that the region where $||v_\pm||$ is of order $1$ also corresponds to that
where the equations of the two theories differ significantly. Outside  this region, the equations and speeds of the two theories are very close. Thus, we conjecture that the physical evolution of the scalar field 
outside the region where $||v_\pm||\sim1$ is trustworthy. We checked that this behavior persists at late times, when the scalar field  grows considerably, as a result of spontaneous scalarization.

\begin{figure}
	\centering
	\includegraphics[width=\columnwidth]{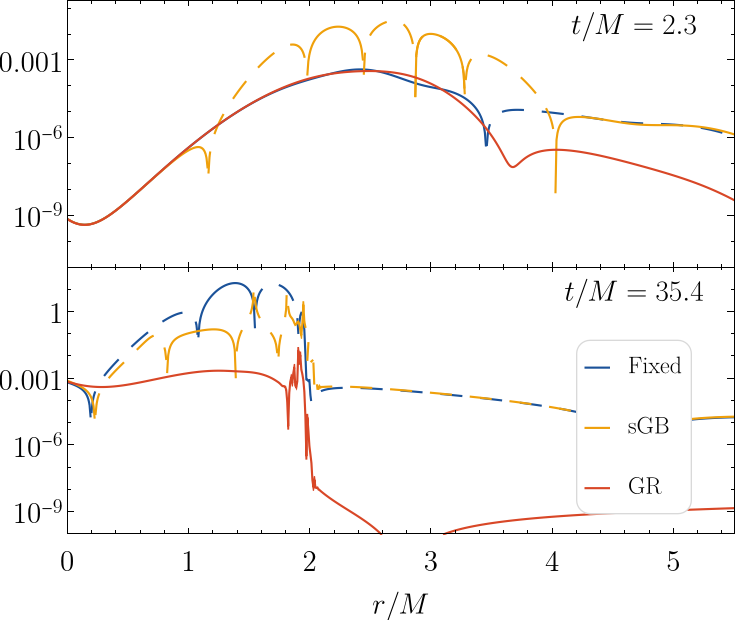}
	\caption{Null convergence condition. 
	Each line corresponds to a radial snapshot of the NCC evaluated in the {\it fixed} system (blue line), in the {\it full} system (yellow line) or with a minimally coupled scalar field (red line). Solid lines correspond to positive NCC, while dashed lines correspond to negative NCC. The top panel shows a snapshot at $t = 2.3 M$ (roughly corresponding to the appearance of an elliptic region in the {\it full} system), while the bottom panel shows the late behavior of the system at $t = 35.4 M$.}\label{fig:NCC}
\end{figure}

Another way to understand how the {\it fixing} works consists of evaluating the NCC. In Fig.~\ref{fig:NCC} we compare the condition for the {\it fixed} system, for the {\it full} one evaluated with quantities in the {\it fixed} system, and for an evolution with equivalent initial data, but with a minimally coupled scalar field (labelled as EV4 in Table~\ref{tab:runs}). We can see that at the formation of the elliptic region (upper panel), the NCC of the {\it fixed} system resembles that of GR 
at small radii, while it almost overlaps with the condition evaluated for the {\it full} system at large radii. This latter trend persists throughout the whole evolution, as one can notice in the lower panel of the figure, evaluated at $t = 35.4M$. The absolute value of the NCC is here expressed in logarithmic scale, we denote the regions where it becomes negative by a dashed line style. These results agree with expectations based on the analysis carried in~\cite{Ripley:2019irj}.

\begin{figure}
	\centering
	\includegraphics[width=\columnwidth]{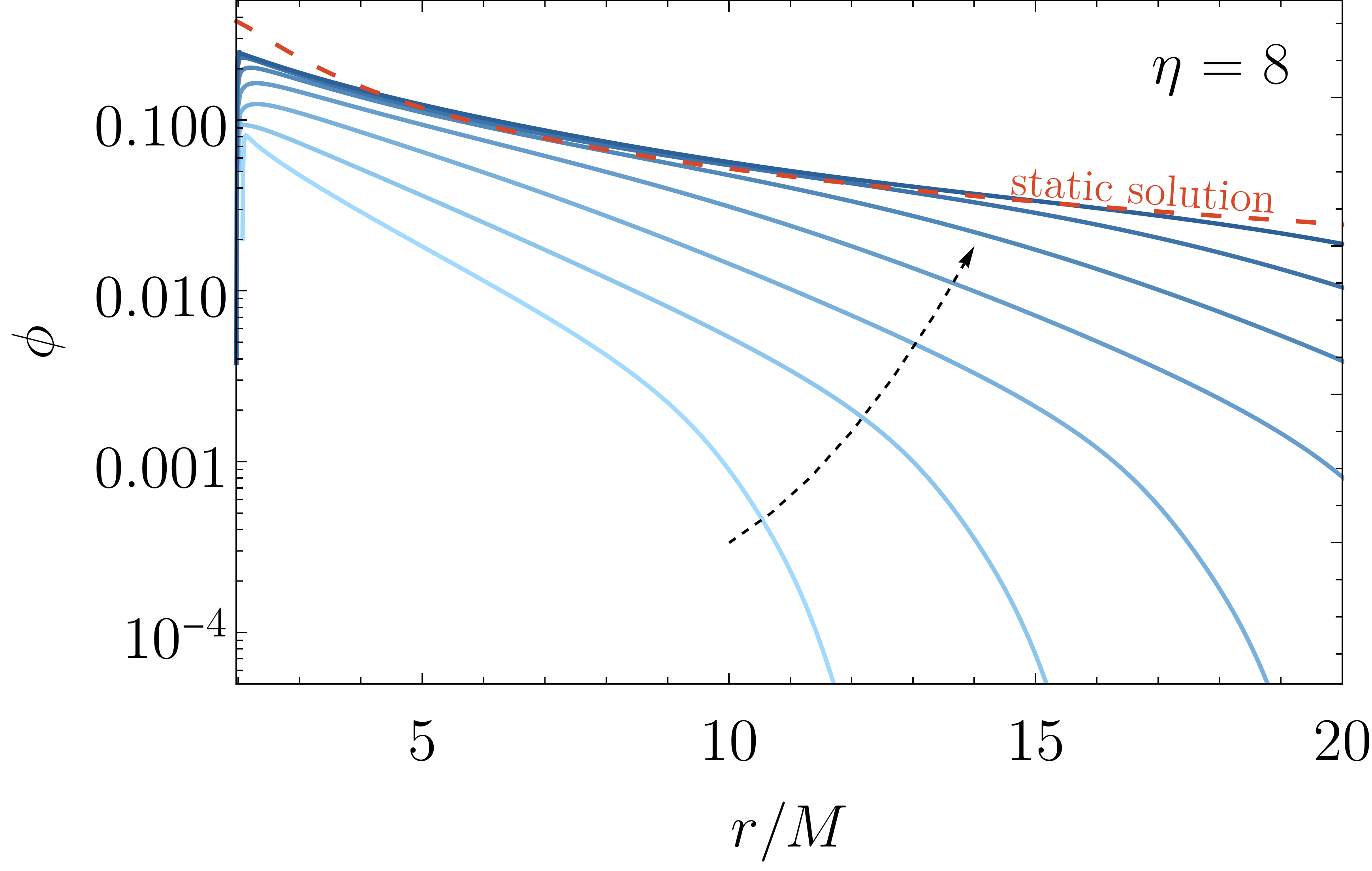}
	\caption{Scalar field profile for the scalar field. The blue solid lines are time snapshots of the radial profile outside  the apparent horizon (after it has formed) as extracted from our simulations, for $\eta = 8$ and $\ze = - 6\eta$. The direction of the evolution is marked by the dotted black arrow, and by the lines getting darker. The red dashed line approached by late-time evolutions of the scalar field is the radial profile in the static case.} \label{fig:phi_e6z6}
\end{figure}

The previous analysis heuristically shows that the numerical evolution of the {\it fixed} theory is reliable, especially outside the apparent horizon (when the latter forms). Thus, we evolved a few cases with different combinations of $\eta$ and $\ze$ to see whether spontaneous scalarization of the final BH that forms from the collapse is achieved. We consider one representative case, summarized in Table~\ref{tab:runs} as EV5. The results for the gravitational collapse of this case are shown in Fig.~\ref{fig:phi_e6z6}. The blue lines indicate time snapshots of the radial profile of the scalar field $\phi$ after the formation of an apparent horizon. After the collapse, the scalar field surrounding the BH settles down to a static solution. We compare this equilibrium configuration with the one obtained by solving the static spherical problem in the {\it full} theory ({\it c.f.}~\cite{Silva:2018qhn}). This static solution can be obtained simply by requiring all the functions to be time-independent, or equivalently by replacing $\partial_{t}\phi=\partial_{t}P=\partial_{t}Q=0,$ in Eqs.~\eqref{eq:Einstein}--\eqref{eq:KleinGordon}. By doing so, the equations are reduced to a set of ordinary differential equations, which can be easily integrated by imposing proper boundary conditions at the horizon of the BH and asymptotic flatness at the outer boundary.
We clearly see that the evolution drives the scalar field to match the static profile. We stress that we obtained similar results with different values of the coupling and different choices of initial data.

A few caveats to discuss are the following. The dynamical evolution of the scalar field does not seem to asymptote well to the static solution in a region very close to the apparent horizon. We conjecture that this behaviour is due to a small portion of the elliptic region of the {\it full} theory  leaking out of the horizon. Therefore, even if we can ``heal'' the character of the field equations in the {\it fixed} theory, we still expect some discrepancy in this region of large gradients and high-frequency energy flows. On the other hand, the scalar field does not fully match the static case at large radii either, but this is simply because we could not run the evolution any longer. At the level of implementation, 
possible improvements include
 using horizon penetrating coordinates rather than polar coordinates and/or  excising the region inside the apparent horizon. 

In order to understand how the fixing behaves, we studied the effect of different values of $\xi$ and $\tau$. We are interested in the limit $\tau\rightarrow 0$ and $\xi\rightarrow0$. However, note that for values of $\xi$ lower than the coupling constant $\eta$, the evolution breaks down at early stages and cannot be followed, as expected from~\cite{Cayuso:2017iqc}. This is an issue that can be
addressed by coupling both parameters in a suitable manner. Otherwise, since the decay rate is $\approx \tau/\xi$, adopting too small a $\xi$ value would yield a stiff system, which needs to be handled with care at the numerical level. In our simulations we could set $\tau$ identically to zero as the results do not change with respect to cases where $\tau/\eta\ll1$.

Notice that a static, radially dependent, solution would induce a discrepancy
between $\bf{u}$ and $\bf{S}$ given by $\approx \xi \nabla^2 \bf{u}$. One can recover the static behavior as $\xi \rightarrow 0$ {\it for the particular fixing employed}.\footnote{Other options,
like having $\xi \bf{u}_{,tt} = - \tau \bf{u}_{,t} + (\bf{u}-\bf{S})$ would be free of this issue, but modes would not propagate away. This shows that the freedom in fixing the equations needs to be properly explored in order to
settle on a fully working choice---see also~\cite{Cayuso:2017iqc}.}
Here, we employ a single value for $\xi$. For completeness, we also checked cases where we adopt five different constants---one for each auxiliary field---and found that we can lower a subset of these constants, while still getting a proper evolution. Namely, by lowering by a factor 10 the constants controlling the equations for $\Sg$, $\Ga_{22}$ and $\Ga_{33}$, we observed that the results obtained do not qualitatively change. 

\begin{figure}
	\centering
	\includegraphics[width=\columnwidth]{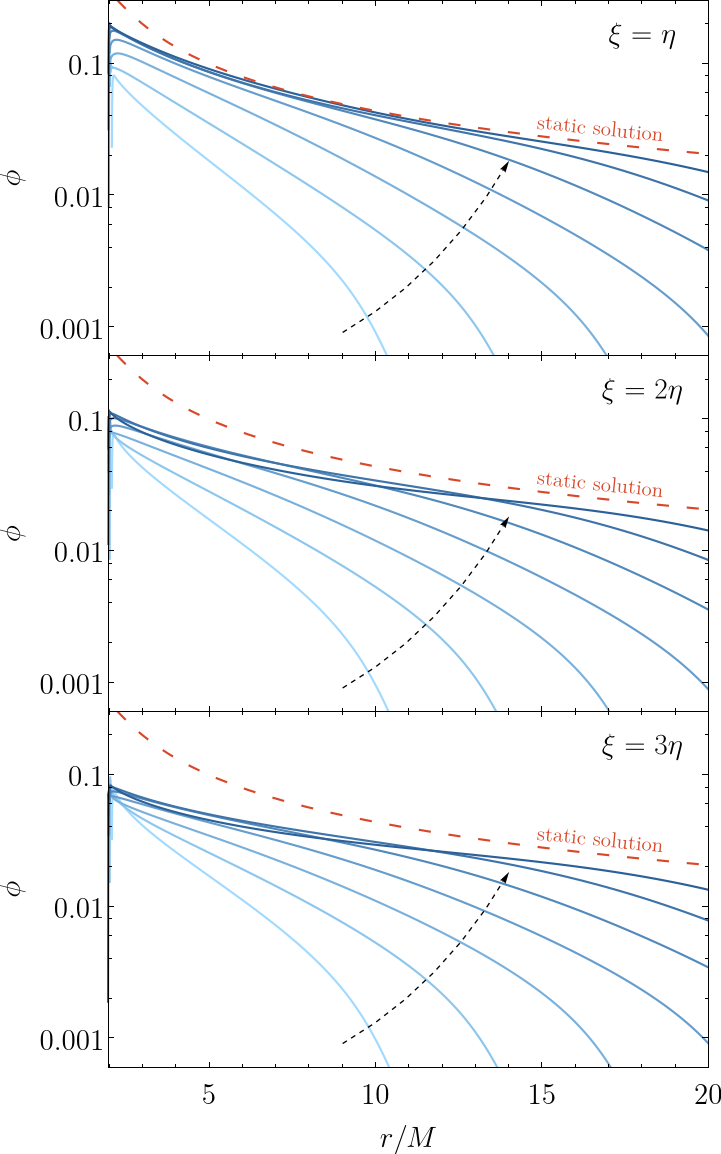}
	\caption{
	{Same as Fig.~\ref{fig:phi_e6z6}, but with for $\eta = 6$, $\ze = -6\eta$ and different values of $\xi$.}}
\label{fig:extrapolation}
\end{figure}

On the other hand, in Fig.~\ref{fig:extrapolation} we show the effects of increasing $\xi$. The three panels are analogous to Fig.~\ref{fig:phi_e6z6}, but with $\eta = 6$ and $\ze = -6\eta$, and each of them assumes an increasing value of $\xi$. The runs correspond to EV3, EV6 and EV7 in Tab.~\ref{tab:runs}, respectively. It is clear that the scalar evolution matches well  the static prediction only for the lowest value of $\xi$. This is not surprising, as large values of  $\xi$ imply that the fixed system's faithfulness to the original one degrades.

\section{Conclusions}\label{sec:conlusions}
In the current work we studied the dynamical collapse and related hair BH formation
in sGB gravity. Such a theory has been the subject of scrutiny in recent
years, which has revealed interesting behaviors for BH physics~\cite{Silva:2017uqg,Doneva:2017bvd,Antoniou:2017acq,Silva:2018qhn,Cunha:2019dwb,Dima:2020yac,Herdeiro:2020wei,Berti:2020kgk}.
However, as it has been discussed elsewhere~\cite{Ripley:2019hxt,Ripley:2019irj,Ripley:2020vpk,Witek:2020uzz,Julie:2020vov}, it also presents
important mathematical challenges in regimes of high couplings and/or short wavelengths. 
Such challenges stem from the equations of motion breaking their hyperbolic character. This makes
it impossible to explore important questions such as the formation and non-linear stability of compact objects, unless further steps are taken to address the the theory's shortcomings.
Further, even in the weak coupling regime, physical effects abound that induce short wavelength 
behavior from long wavelength, {\it e.g.}, focusing and collapse. For instance, a gravitational (or
scalar) wave passing by a BH could be focused and excite significantly short wavelengths, 
potentially breaking (local) hyperbolicity. Unless a BH generically
shields such regions away from physical observers, one could argue that
extensions of GR that present these shortcomings would have limited physical relevance. It behooves
theorists to try and devise ways to explore such question.

Here, we adopted a strategy dubbed as {\it fixing-the-equations}, which introduces further auxiliary
fields, constrained within some given scale, to force a suitably modified system to
agree with the original equations of motion, while at the same time controlling the behavior at shorter wavelengths~\cite{Cayuso:2017iqc}.
This approach, to a certain extent, recovers some of the higher energy degrees of freedom, which
in an effective field theory sense are integrated out and whose influence results in corrections
to the Einstein-Hilbert action. Controlling the high frequency modes of the solution, this approach
eliminates (or significantly reduces) the mathematical pathologies of the original system, thus
allowing one to push evolutions further. Note that the {\it fixing} approach 
breaks Lorentz symmetry, in particular through the $\tau$-term in Eq.~\eqref{eq:IS_Aux} which sets a scale---in time and wavelength---below which such breaking is especially evident. The presence of this term is in agreement with the fact that the UV completion of the theory has to be non-standard~\cite{Herrero-Valea:2021dry}.

In our particular application, the {\it fixing} allowed us
to resolve the collapse to a scalarized BH. Remarkably, this dynamically formed BH agrees with static solutions of the original unfixed theory. We stress that this was not necessarily expected, as our method does introduce
modifications in the theory. As such, the agreement of the solutions found is highly non-trivial and indicates 
that the fixed system successfully bridges between stages where/when
the original system is well behaved. 
We also stress that we find regimes where hyperbolicity breaks {\em outside} the apparent horizon for sGB as already noted in previous works~\cite{Ripley:2019hxt,Ripley:2019hxt,Ripley:2020vpk,East:2020hgw,East:2021bqk}. Two comments are in order here: (i) that sGB shows fails to be hyperbolic outside
the horizon implies one can not assume that a horizon will hide such failure from outside observers.
(ii) this failure should not happen in a proper UV completion of the theory, and the fixed version
allows to by pass such issue.
Last, we note that
the fixed system yields a solution that
approaches the one expected in the static regime.
This is analogous to the behavior observed in an effective-field-theory-derived gravitational
theory where BH accretion was studied and the late time solution agreed with the static
one~\cite{Cayuso:2020lca}.
If this observed robustness could be extrapolated to more general solutions and
other extensions to GR, it would lend strong support to the physical relevance of such solutions
and conclusions drawn from perturbative studies.
Furthermore, this agreement supports the idea that the approach employed can be regarded as 
furnishing a weak completion of the original theory. 

\acknowledgments
We thank R.~Cayuso, G.~Dideron, A.~Dima, W.~East, G.~Lara, C.~Palenzuela and A.~Tolley for discussions related to this work.
We acknowledge financial support provided under the European Union's H2020 ERC Consolidator Grant ``GRavity from Astrophysical to Microscopic Scales'' grant agreement no.~GRAMS-815673. This work was supported by the EU Horizon 2020 Research and Innovation Program under the Marie Sklodowska-Curie Grant Agreement No.~101007855. Also, this research was supported in part by CIFAR, NSERC through a Discovery grant, and by Perimeter Institute for Theoretical Physics. Research at Perimeter Institute is supported by the Government of Canada and by the Province of Ontario through the Ministry of Research, Innovation and Science

\appendix

\section{Code validation}

\begin{figure}
	\centering
	\includegraphics[width=\columnwidth]{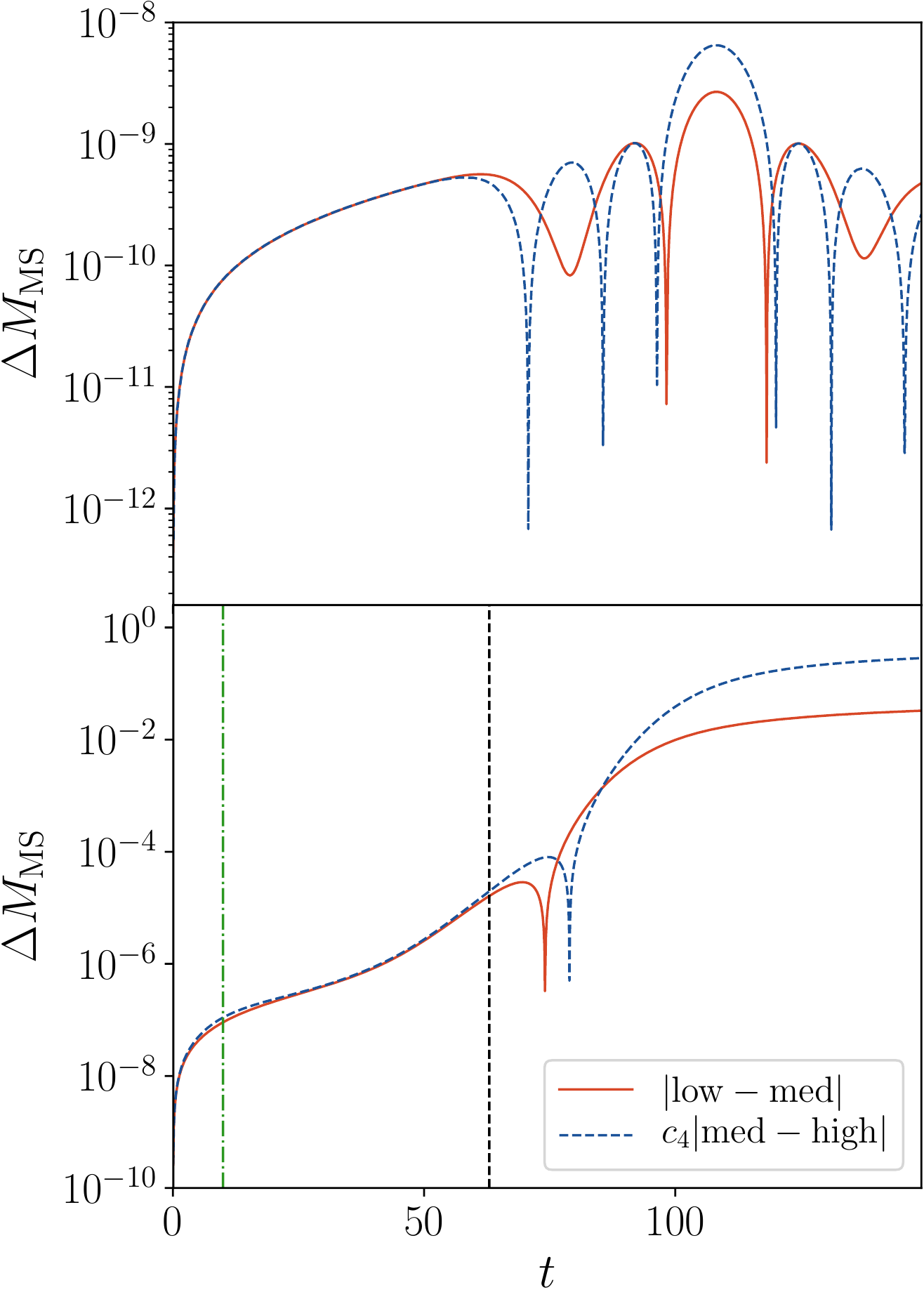}
	\caption{Convergence for the Misner-Sharp mass for the cases EV8 and EV3 respectively. We display in both panels the absolute difference between the low and medium resolutions (solid red line) and the medium and high resolutions (dashed blue line). The latter is rescaled by a factor $c_{4} = 16$ in both cases, showing fourth order convergence. The green dot-dashed line denotes the time when the elliptic region in the {\it full} theory appears, and the dashed black line indicates when the BH is formed.} \label{fig:conv}
\end{figure}

Here we present the details about the convergence tests performed to validate our simulations. We evolve the \textit{fixed} theory considering two representatives cases: (i) EV8 (see Table~\ref{tab:runs}), where a small pulse just grows as it approaches to the origin and then disperses to infinity, and (ii) EV3, where the pulse collapses to form an apparent horizon, with a non-trivial scalar profile outside. Our simulations are evolved by using three different resolutions $\{\De r_{\text{\rm{low}}},\De r_{\text{\rm{med}}},\De r_{\text{\rm{high}}}\}=\{0.0625,0.03125,0.015625\}$. We compute the convergence factor $c_{p}$ and the order of convergence as follows
\begin{equation}
	c_{p} = \frac{\left|(\Delta r_{\text{\rm{low}}})^{p}-(\Delta r_{\text{\rm{med}}})^{p}\right|}{|(\Delta r_{\text{\rm{med}}})^{p}-(\Delta r_{\text{\rm{high}}})^{p}|}\,.
\end{equation}
We use the Misner-Sharp mass~\eqref{MSmass} to compute  $c_{p}$ and $p$, as shown in Fig.~\ref{fig:conv}. In the top panel, we show the convergence for EV8, finding a fourth order convergence as expected. In the lower panel the convergence for EV3 is displayed. In this case,  until the collapse we find fourth-order convergence, but after the formation of the apparent horizon the convergence is somewhat degraded. This is expected due to the large gradients in the collapse front. 

\section{Source of the fixed equations}\label{app:source}

In this Appendix we show the explicit form of the source terms $\mbf{S}$ appearing in equation~\eqref{eq:IS_Aux}

\begin{widetext}

\begin{subequations}
\begin{equation}
\begin{split}
    S_\Sg  =  f'(\phi_0) \Bigg[ & \frac{16 \ee^{-4 B} \dl_rA \dl_rB+2\left(\ee^{2 B}-\ee^{-2 B}\right) \left(P^2-Q^2\right)}{r^2}-\frac{8 \left(\ee^{2 B}-\ee^{-2 B}\right)
   \left(\dl_rA-\dl_rB\right)}{r^3} \left. -\ee^{-4B}\De^2+\frac{8 y \Gamma _{33}}{r^4}\right] \,,
\end{split}
\end{equation}  
\begin{equation}
\begin{split}   
   S_{11} = \ee^{2 A-2 B}f'(\phi_0) \Bigg[  \frac{4 \left(\left(3 \ee^{-2 B}-1\right) Q \dl_rB+y \dl_rQ\right)}{r^2}   - \left. \frac{y P 
   \De}{r}\right]  +\frac{4 y \, \ee^{2 A-2 B} f''(\phi_0)  Q^2 }{r^2} \,,
\end{split}
\end{equation} 
\begin{equation}
\begin{split}   
   S_{12} =-\ee^{A-B} f'(\phi_0) \Bigg[ &\frac{4y  \left(P \dl_rB-\dl_rP\right)}{r^2} +\frac{\left(1-3 \ee^{-2 B}\right) Q \De}{r}\Bigg]+\frac{4y f''(\phi_0)  P Q \ee^{A-B}}{r^2}  \,,
\end{split}
\end{equation}
\begin{equation}
\begin{split}   
   S_{22} = f'(\phi_0) \Bigg[ &\frac{8 \ee^{-2 B} Q \dl_rA-4y \left(Q \dl_rB-\dl_rQ+\ee^{2 B} \Sigma  \right)}{r^2} \left. -\frac{y P \De}{r}+\frac{8 y Q}{r^3}\right]+\frac{4y f''(\phi_0)  P^2}{r^2} \,,
\end{split}
\end{equation} 
\begin{equation}
\begin{split}   
   S_{33} = \ee^{-4 B} f'(\phi_0) \Bigg[ & r \bigg(\!-4 \dl_rB \left(Q \dl_rA-\dl_rQ+\ee^{2 B} \Sigma  \right)+4 \dl_rA \dl_rQ-4 Q
   \left(\dl_rB\right)^2 +P^2 Q-Q^3 \bigg) \\
   &    - r^2  \left(P \dl_rA-P \dl_rB+2 \dl_rP\right) \De \left. - 4 Q
   \left(\dl_rA-3 \dl_rB\right) + \frac{1}{2} Q r^3 \De^2+\frac{4\ee^{2B}\Gamma _{33} Q}{r}\right] \\
   +2e^{-4 B} f''(\phi_0) \bigg[ & 2 r
   \left(Q^2 \dl_rA+P^2 \dl_rB\right)- P Q r^2 \De \bigg] \,,
\end{split}
\end{equation}
\end{subequations}

\end{widetext}

with $y = \left(1-\ee^{-2 B}\right)$ and $\De =  \left(P Q+2 \ee^{B-A} \Gamma _{12}\right)$.

\bibliography{bibnote}

\end{document}